# SOFIA Polarization Spectrum of Three Star-forming Clouds

Erin G. Cox[1,8], Giles Novak[1,2], David T. Chuss[3], Dennis Lee[1,2], Marc Berthoud[1,4], Kaitlyn Karpovich[3], Joseph M. Michail[1,2,5,9], Zhi-Yun Li[6], and Peter C. Ashton[7]

[1] Center for Interdisciplinary Exploration and Research in Astronomy (CIERA), Northwestern University, 1800 Sherman Avenue, Evanston, IL 60208, USA; erin.cox@northwestern.edu
[2] Department of Physics & Astronomy, Northwestern University, 2145 Sheridan Road, Evanston, IL 60208, USA
[3] Department of Physics, Villanova University, 800 E. Lancaster Ave., Villanova, PA 19085, USA
[4] Engineering + Technical Support Group, University of Chicago, Chicago, IL 60637, USA
[5] Center for Astrophysics | Harvard & Smithsonian, 60 Garden St., Cambridge, MA 02138, USA
[6] Department of Astronomy, University of Virginia, P. O. Box 400325, 530 McCormick Road, Charlottesville, VA 22904-4325, USA
[7] SRI International, 333 Ravenswood Ave., Menlo Park, CA 94025, USA

*Received 2024 July 24; revised 2024 December 17; accepted 2024 December 29; published 2025 March 4*

## Abstract

The dust emission polarization spectrum—how the polarization percentage changes with wavelength—serves as a probe of dust grain properties in star-forming regions. In this paper, we present 89–214 $\mu$m polarization spectrum measurements obtained from SOFIA/HAWC+ for three star-forming clouds: OMC1, M17, and W3. We find that all three clouds have an overall decreasing polarization percentage with increasing wavelength (i.e., a "falling polarization spectrum"). We use SOFIA and Herschel data to create column density and temperature maps for each cloud. We fit for the slope of the polarization spectrum at each sky position in each cloud, and using the Pearson $r$ coefficient, we probe each cloud for possible correlations of slope with column density and slope with temperature. We also create plots of slope versus column density and slope versus temperature for each cloud. For the case of OMC1, our results are consistent with those presented by J. Michail et al., who carried out a similar analysis for that cloud. Our plots of polarization spectrum slope versus column density reveal that for each cloud there exists a critical column density below which a falling polarization spectrum is not observed. For these more diffuse sight lines, the polarization spectrum is instead flat or slightly rising. This finding is consistent with a hypothesis presented 25 yr ago in a paper led by R. Hildebrand based on Kuiper Airborne Observatory data. This hypothesis is that regions shielded from near-IR radiation are required to produce a sharply falling polarization spectrum.

*Unified Astronomy Thesaurus concepts:* Star formation (1569); Star forming regions (1565); Far infrared astronomy (529); Polarimetry (1278); Interstellar magnetic fields (845); Protostars (1302); Young stellar objects (1834)

## 1. Introduction

Fossil magnetic fields in the Milky Way galaxy thread star-forming clouds and are thought to be a crucial aspect in the collapse of the natal clouds to form protostars (e.g., C. Federrath & R. S. Klessen 2012). These fields can directly influence where stars form, guiding the molecular gas and causing dense regions to accumulate and eventually give rise to young stars. They can also provide magnetic support to the natal clouds, thus hindering gravitational collapse on short timescales.

Measurement of dust polarization is the most common method to observationally constrain magnetic fields in star-forming regions (e.g., K. Pattle et al. 2023). Detecting these magnetic fields can be difficult, as polarized emission is often only a small fraction of the total dust emission and thus requires quite sensitive instruments and long integration times. Dust polarization maps only provide one aspect of the magnetic field, the plane-of-sky morphology. Both the strength of the field and its morphology are thought to be important in constraining how the field affects star formation. This has motivated the development of statistical techniques that use dust polarization maps to infer the magnetic field strength (e.g., via the DCF method; L. Davis 1951; S. Chandrasekhar & E. Fermi 1953) and to test dust grain alignment mechanisms.

These dust polarization observations rely on an understanding of the mechanism by which the dust grains align relative to the local magnetic field. The most common paradigm is that of radiative alignment torques (RATs; e.g., A. Z. Dolginov & I. G. Mitrofanov 1976; B. T. Draine & J. C. Weingartner 1997; A. Lazarian 2007; A. Lazarian & T. Hoang 2007; B. G. Andersson et al. 2015). RAT theory relies on optical/near-IR radiation to torque a spinning, elongated dust grain such that its short axis aligns with the local magnetic field. The thermal emission of these aligned grains is thus polarized in the direction orthogonal to the magnetic field. Constraining the dust emission polarization spectrum—how the polarization percentage changes with wavelength—provides a method for testing RAT theory.

R. H. Hildebrand et al. (1999) used the Kuiper Airborne Observatory (KAO) to carry out the first observations of the far-IR polarization spectrum for star-forming regions. For OMC-1 and M17, they compared the KAO polarimetry results with ground-based polarimetry at a longer wavelength (350 $\mu$m), finding that the polarization percentage falls with wavelength—i.e., negatively sloped polarization spectra. After showing that the intrinsic polarization spectrum of a homogeneous grain population is expected to be quite flat,

---

[8] NSF MPS-Ascend Postdoctoral Fellow.
[9] NSF Astronomy and Astrophysics Postdoctoral Fellow.

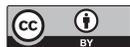

R. H. Hildebrand et al. (1999) argued that their observations were best explained by invoking a heterogeneous cloud where typical sight lines contain separate warmer and cooler regions, with the grains in the cooler regions being more poorly aligned than those in the warmer regions. J. M. Michail et al. (2021) explored the far-IR polarization spectrum of OMC-1 using SOFIA/HAWC+ (D. A. Harper et al. 2018) multiwavelength data that were first reported by D. T. Chuss et al. (2019). J. M. Michail et al. (2021) again found negatively sloped spectra, and they invoked the same hypothesis advanced by R. H. Hildebrand et al. (1999) to explain their results, coining the term heterogeneous cloud effect (HCE) to refer to the hypothesis. R. H. Hildebrand et al. (1999) noted that the HCE is compatible with RATs because grains that are preferentially exposed to optical/near-IR radiation will be both better aligned and warmer than grains that are less exposed to such radiation.

In order to perform additional tests of RATs and dust grain physics, we here extend the study of SOFIA/HAWC+ polarization spectra to a sample of three high-mass star-forming regions: OMC-1, M17-SW, and W3 Main. These targets were chosen because they are relatively nearby and because extensive multiwavelength HAWC+ polarization data are available for all three targets in the SOFIA archive. All three contain high-mass young stellar objects and/or high-mass protostars that are embedded in far-IR dust emission peaks (e.g., M. Kassis et al. 2002; L. Testi et al. 2010; A. Rivera-Ingraham et al. 2013). The distances vary from ~390 pc for OMC-1 (M. Kounkel et al. 2017), which is the closest site of high-mass star formation, to ~1700 and ~2200 pc for M17-SW and W3 Main, respectively (M. A. Kuhn et al. 2019; F. Navarete et al. 2019). For all three star-forming regions, the general distribution of the 214 $\mu$m emission is dominated by two emission peaks, as can be seen in the right column of Figure 1. The peaks are separated by approximately 0.17, 0.56, and 0.72 pc for OMC-1, M17-SW, and W3 Main, respectively. Estimated gas masses for our three targets range from ~290 $M_\odot$ for OMC-1 (E. S. Wirström et al. 2006) to ~4500 and ~4100 $M_\odot$ for M17-SW and W3 Main, respectively (M. Hayashi et al. 1989; M. P. Hobson et al. 1993).[10]

In their study of OMC-1 HAWC+ data, J. M. Michail et al. (2021) separately correlate the slope of the far-IR polarization spectrum with each of two environmental parameters: column density and temperature. Their computations of these environmental parameters rely on archival OMC-1 maps from six telescopes at 11 wavelengths, but such rich data sets are not available for our other two targets. In extending the analysis of J. M. Michail et al. (2021) to these other clouds using more limited data sets, we check the fidelity of our analysis by re-creating the previous work with a relevant subset of the data, namely, the HAWC+ polarimetric and photometric data and Herschel[11] SPIRE 350 $\mu$m photometry.

We limit our study of the polarization spectrum to only include the HAWC+ wavelengths instead of combining with submillimeter data. Some previous observations have found that the polarization spectrum rises again at these longer wavelengths (e.g., J. E. Vaillancourt et al. 2008; J. E. Vaillancourt & B. C. Matthews 2012), while other studies found flatter polarization spectra (e.g., N. N. Gandilo et al. 2016; J. A. Shariff et al. 2019). The combination of polarimetric data from multiple telescopes can be affected by biases in the polarization fraction at a given location if different methods are used for spatial referencing (e.g., different chopper throws or different separations between target and reference regions). These biases can be seen in Figure 4 of L. M. Fissel et al. (2016). Consideration of such biases is beyond the scope of the present paper.

In Section 2, we discuss the Herschel and HAWC+ data sets, as well as the data analysis procedures we used. In Section 3, we present the resulting HAWC+ polarization maps, the derived column density and temperature maps, the computed HAWC+ polarization spectra, and the results of our study of correlations between various quantities. In Section 4, we discuss the implications of our results.

## 2. Data and Observations

### 2.1. SOFIA Data

We used SOFIA/HAWC+ polarimetric observations of three star-forming clouds to determine their far-IR polarization spectrum. OMC-1, M17-SW, and W3 Main (hereafter OMC-1, M17, and W3) were observed in all four bands of HAWC+ (53, 89, 154, and 214 $\mu$m). OMC-1 was observed on 2016 December 3, 2017 February 19, and 2017 October 24, with an observing time of 5869, 3906, 757, and 1035 s at 53, 89, 154, and 214 $\mu$m, respectively. The observations of M17 were taken from 2018 September 20 to 2018 September 21 and were observed for 908 s at 53 $\mu$m, 2333 s at 89 $\mu$m, 942 s at 154 $\mu$m, and 710 s at 214 $\mu$m. W3 was observed on 2016 December 1, 2016 December 3, 2016 December 6, and 2017 October 17 with an observing time of 5315 s at 53 $\mu$m, 1527 s at 89 $\mu$m, 1418 s at 154 $\mu$m, and 2649 s at 214 $\mu$m. For our analysis, however, we opt for larger spatial coverage and omit the 53 $\mu$m data owing to its small field of view in most targets. These observations utilized a standard matched-chop-nod method (R. H. Hildebrand et al. 2000), where the chopping frequency was 10.2 Hz and the chop angle and chop throw differed for each cloud and wavelength. The choices of the angle and throw were made such that the flux in the reference beam was minimized. Each observation was taken using a four-point dithering block, with the offset dependent on source and wavelength. A complete dither set forms a square on the sky. OMC-1 polarization data were first published in D. T. Chuss et al. (2019) at the nominal resolution for each band: 53 $\mu$m, 5″; 89 $\mu$m, 8″; 154 $\mu$m, 14″; and 214 $\mu$m, 19″. T. D. Hoang et al. (2022) presented the 154 $\mu$m data for M17, but this is the first time that the other polarization maps of M17 and W3 have been shown.

To reduce the HAWC+ data, we used the data reduction pipeline described in F. P. Santos et al. (2019). In this pipeline, the data are first demodulated and any flagged or bad data are discarded. This first step also takes into account the fact that the observations are taken in chop-nod mode. Then, we calibrate any variations in gain between pixels using our flat-field data. This flat-fielding removes data from both dead and noisy pixels. We then create Stokes $I$, $Q$, and $U$ maps for each independent pointing by first subtracting signals that are reflected by the polarizer from those that are transmitted. Stokes $I$ is determined using the combined measured flux at each nod position. We use standard models to apply an atmospheric correction to each map. These results from each pointing are then combined to create final Stokes $I$, $Q$, and $U$ maps.

---

[10] The mass estimates have been corrected to reflect updated distance estimates given in this section.

[11] Herschel is an ESA space observatory with science instruments provided by European-led Principal Investigator consortia and with important participation from NASA.

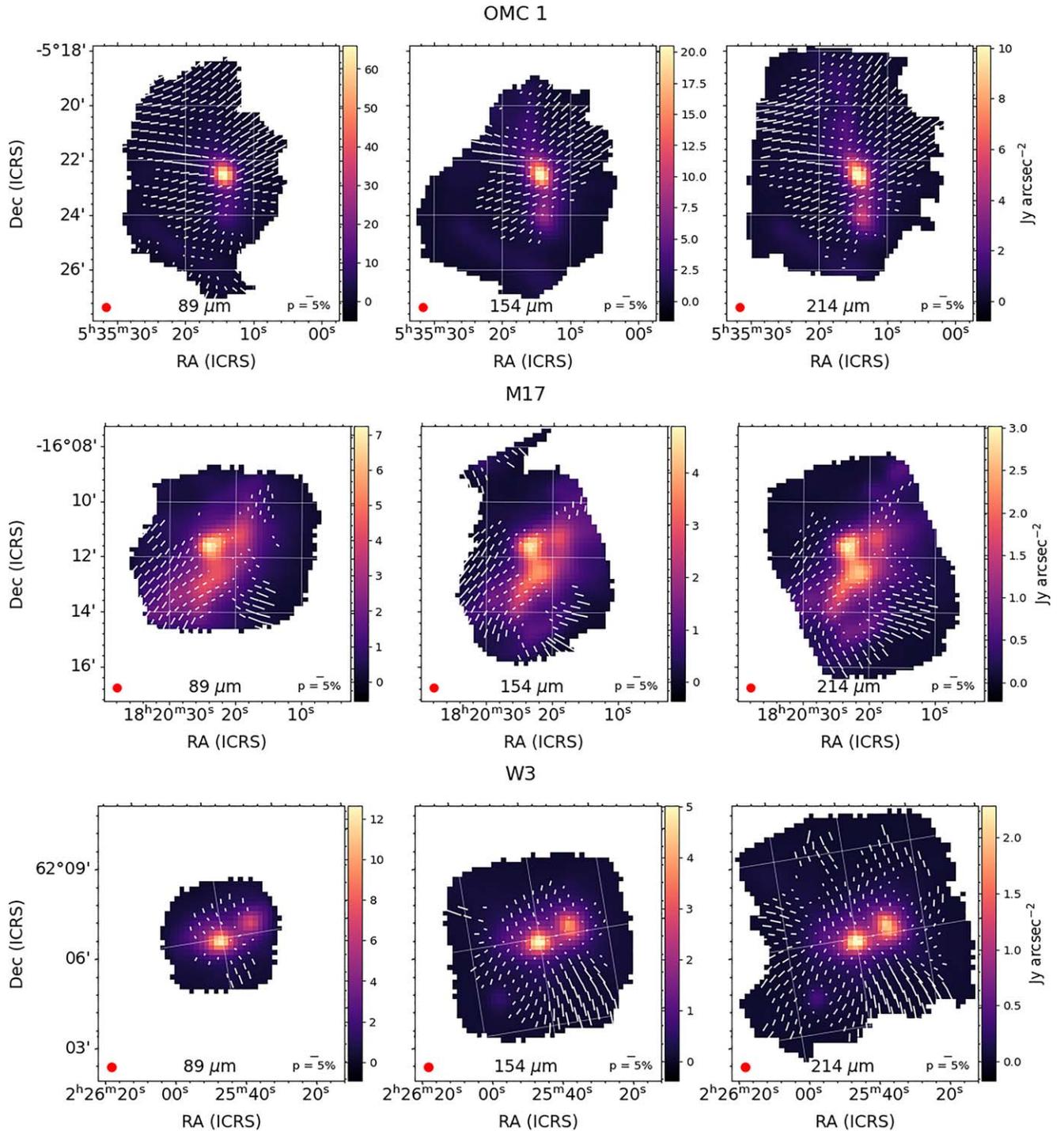

**Figure 1.** Inferred magnetic field (rotated 90° polarization vectors) of OMC-1 (top), M17 (middle), and W3 (bottom) shown in the white line segments. Independent (i.e., one per beam) measurements are shown. The polarization percentage of each is proportional to the length of the line, with 5% shown in the lower-right corner. These images are all smoothed to the Herschel 350 μm resolution of ~25″, shown by the red beam in the lower-left corner of each panel. The left column is 89 μm data, the middle column is 154 μm, and the right column is 214 μm. The background image is the total intensity emission for each band. The data shown here only encompass data cuts 1–3 described in Section 2.1.

We perform a $\chi^2$ analysis on our data to check for systematic error in the Stokes $Q$ and $U$ maps. This is done by first dividing the data into bins and computing Stokes $Q$ and $U$, as well as their corresponding uncertainties. We compare the scatter in the maps to the expected value based on the uncertainties and then inflate the error bars accordingly (see, e.g., G. Novak 2011; N. L. Chapman et al. 2013). As described in E. G. Cox et al. (2022), we do this check in a robust way, accounting for differences in pixel intensity (i.e., each pixel's errors are inflated corresponding to the intensity of that pixel). Though this robust form of the $\chi^2$ analysis was conceived to account for low flux targets, we use this error inflation exclusively in our Stokes $Q$ and $U$ maps (i.e., not for Stokes $I$) to better estimate our polarization uncertainties.

To perform our polarization analysis, we first create a data set that can be comparable across wavelengths. We smooth each pipeline-reduced SOFIA observation to the Herschel 350 μm resolution of ~25″ using astropy's (Astropy Collaboration et al. 2013; A. M. Price-Whelan et al. 2018; Astropy Collaboration et al. 2022) implementation of a Gaussian kernel, whose size is found using $\text{FWHM}_{\text{smooth}} = \sqrt{(25'')^2 - (\text{FWHM}_\lambda)^2}$. This is done to each Stokes parameter and the errors associated with these maps. We then reproject these smoothed data onto a common pixel size of 10″ (i.e., the pixel size of the Herschel 350 μm maps) using a flux-conserving algorithm (Astropy Collaboration et al. 2013; A. M. Price-Whelan et al. 2018; Astropy Collaboration et al. 2022). With these data, we then compute the polarization intensity and uncertainty using

$$P = \sqrt{Q^2 + U^2}, \quad (1)$$

$$\sigma_P = \frac{\sqrt{(Q\sigma_Q)^2 + (U\sigma_U)^2}}{P}, \quad (2)$$

where $Q$ and $U$ are the linear Stokes parameters and their associated uncertainties are $\sigma_Q$ and $\sigma_U$. To compute the fractional polarization, we use the formula

$$p_{\text{frac}} = \frac{P}{I}. \quad (3)$$

We then debias this using the most probable estimator (e.g., J. F. C. Wardle & P. P. Kronberg 1974),

$$p = \sqrt{p_{\text{frac}}^2 - \sigma_p^2}, \quad (4)$$

where $\sigma_p$ is the uncertainty in $p_{\text{frac}}$ and is found using

$$\sigma_p = p_{\text{frac}} \sqrt{\left(\frac{\sigma_P}{P}\right)^2 + \left(\frac{\sigma_I}{I}\right)^2}, \quad (5)$$

where $I$ is the total intensity, $\sigma_I$ is the uncertainty in total intensity, $P$ is the polarized intensity, and $\sigma_P$ is the uncertainty in polarized intensity. We assume that the errors in $Q$ and $U$ are uncorrelated. The difference between the debiased and non-debiased polarization percentage is typically ≲1%. In this analysis, we exclusively use the debiased polarization percentage $p$. Finally, we compute the polarization angle using

$$\theta = \frac{1}{2} \arctan\left(\frac{U}{Q}\right). \quad (6)$$

We take care to use the correct quadrant when calculating $\theta$ since the signs for $Q$ and $U$ can determine the polarization angle.

In our analysis across the three wavelengths, we make various cuts to the data to ensure the robustness of our measurements. We first make reference beam corrections to the data (described in more detail in Section 2.2). We then make the following SOFIA polarization data cuts: (1) polarization percentage is at least a $3\sigma$ detection, (2) the polarization percentage is ≲50%, (3) Stokes $I$ is at least 1% of its peak value, and (4) Stokes $I$ is at least a $10\sigma$ detection. In instances where we are comparing the polarization between wavelengths we require that differences in polarization angle between wavelengths are no larger than 15°. This reduces the possibility that our polarization spectrum may be affected by changes in magnetic field direction along the line of sight (J. E. Vaillancourt 2002).

### 2.2. Reference Beam Contamination

To understand the effect of far-IR flux in the HAWC+ reference beams, we used both SPIRE and PACS data. For this reference beam mitigation and our column density and temperature fits, we used Photodetector Array Camera (PACS)[12] 70, 100, and 160 μm data and Spectral and Photometric Imaging Receiver (SPIRE)[13] 250 and 350 μm data. We performed graybody fits (Equation (7)) to compute synthetic fluxes for each SOFIA wavelength. Due to the nature of the chop-nod observations, for each cloud presented in this paper it is possible to have a reference beam region with potentially high levels of polarized intensity. Since the level of polarization is unknown in these regions, we use the formalism laid out in G. Novak et al. (1997) and the results of D. T. Chuss et al. (2019) assuming a maximum reference beam polarization of 10%. This provides a maximum estimated systematic error in polarization angle. If this angle exceeds 10°, the data point is rejected.

## 3. Results

### 3.1. Inferred Magnetic Field

Figure 1 shows the inferred magnetic field (i.e., the polarization vectors are rotated by 90°) morphology in all three star-forming clouds at each observing wavelength. Each map has been smoothed to a 25″ resolution, shown as the red beam in the lower-left corner of each panel, and reprojected onto 10″ pixels. Magnetic field vectors are shown at the beam scale (~25″) such that they represent independent measurements. The magnetic field vectors are scaled for polarization percentage. The background image in each panel is the total intensity (Stokes $I$) for each cloud.

OMC-1 (top panels of Figure 1) exhibits a very uniform magnetic field morphology at all three wavelengths that is consistent across wavelengths. We see the typical pinching, or hourglass, signature toward the central source, indicating that the field is tracing the collapse (D. A. Schleuning 1998). In the northeast region of the 89 and 214 μm maps we see a smooth transition in the morphology of the field in the direction of northwest to southeast. This is consistent with the nonsmoothed maps of OMC-1 presented in D. T. Chuss et al. (2019). At all wavelengths we see the Orion BN/KL region as the brightest source in the total intensity. As the wavelength increases, we also see Orion South become visible just below BN/KL.

M17 (middle panels of Figure 1) shows less of a pinch and more of a dramatic change in angle where the plane-of-sky field is moving across the compression front (see J. L. Dotson 1996; L. Zeng et al. 2013). The polarization data in this cloud do not trace the central, dense region owing to the cuts we imposed on the data. We see that the inferred field is mostly consistent across the three wavelengths. T. D. Hoang et al. (2022) present the 154 μm SOFIA polarization data in M17 at the nominal resolution of ~14″. Their map differs from ours owing to the smoothing and restrictive cuts in our data. Nonetheless, we recover a similar overall magnetic field morphology in M17. In the total emission maps of M17, we see the southern central target becoming more prominent as the wavelength increases.

---

[12] PACS observing labels 1342204845, 1342204846, 1342250631, 1342250632, 1342206052, and 1342206053.
[13] SPIRE observing labels 1342204845, 1342204846, 1342218997, 1342218998, 1342241159, 1342241160, 1342189702, 1342216019, 1342216020, 1342239796, 1342239797, 1342250631, 1342250632, 1342184386, and 1342239930.

The bottom panels of Figure 1 show the inferred magnetic field of W3. In the left column (89 μm), we see possible evidence for a pinched morphology; however, it is not as obvious as in OMC-1. Both the 154 μm (middle panel) and 214 μm (right panel) maps show consistency in their field structures to the 89 μm, with more obvious pinching toward the central dense regions. The 214 μm map recovers a larger area of polarization data and exhibits a more complex morphology than what is seen in the shorter wavelengths. We also see that the western source becomes more prominent in total intensity as the wavelength increases.

The KAO observed all three of these clouds and detected far-IR polarization in all of them (J. L. Dotson et al. 2000). The polarization morphology of OMC-1 at 100 μm was originally presented in D. A. Schleuning (1998) and has a resolution of 35″. J. L. Dotson et al. (2000) reanalyzed these data and presented polarization vectors that have very similar morphology to the ones seen from SOFIA/HAWC+. M17 was observed at 60 μm with 22″ resolution (J. L. Dotson et al. 2000) and 100 μm (J. L. Dotson 1996) and exhibits an inferred magnetic field direction that is roughly the same as our results; however, these observations also probe the peak intensity region, whereas ours do not. The W3 results at 60 and 100 μm from KAO (J. L. Dotson et al. 2000; D. A. Schleuning et al. 2000) show a magnetic field morphology that is quite similar to the one presented here. Additionally, OMC-1 and M17 have been observed by the CSO/SHARP instrument at 350 and 450 μm (J. E. Vaillancourt et al. 2008; L. Zeng et al. 2013).

### 3.2. Column Density and Temperature Fits

To put our polarization data into context, we utilized Herschel 350 μm with the SOFIA total intensity observations to model temperature ($T$) and column density ($N$) in each cloud presented in Figure 2. These maps were created via graybody fits to four different wavelengths, our SOFIA/HAWC+ data plus Herschel 350 μm observations, for each target. As described in Section 2.1, HAWC+ intensity maps were all smoothed and reprojected onto the Herschel 350 μm grid prior to fitting. We used scipycurvefit to do pixel-by-pixel fits to find the optimal column density and temperature in our targets using the following equation:

$$F_\nu = B_\nu(T)\left(1 - e^{-\tau_0\left(\frac{\nu}{\nu_0}\right)^\beta}\right). \quad (7)$$

Here $F_\nu$ is the observed flux per unit solid angle, $\nu$ is the frequency of the map, $B_\nu$ is the Planck equation (which is temperature dependent), and $\tau_0$ is the optical depth at the reference wavelength ($\lambda_0 = 154\,\mu$m for each cloud). In our fits, we chose a constant value of $\beta = 1.6$ (Planck Collaboration et al. 2014) to minimize the number of fitted parameters. The longest wavelength used in these fits was 350 μm, which is not necessarily long enough to constrain the cold dust in the clouds. We converted $\tau_0$ to molecular hydrogen column density ($N$) using the standard relationship $\tau_0 = \kappa\mu m_H N$, where we used $\mu = 2.8$ as the mean molecular weight, $m_H$ is the mass of a hydrogen atom, and $\kappa$ is the dust opacity per unit gas mass at our reference wavelength (0.22 cm$^2$ g$^{-1}$ using the assumptions in R. H. Hildebrand 1983). We compared our fitted values of $N$ and $T$ in each target with published values to ensure that our fits were reasonable.

D. T. Chuss et al. (2019) did extensive fitting in OMC-1 to constrain the column density and temperature using measurements at 11 wavelengths. Our column density results for OMC-1 deviate by a factor of ~4, and our temperature fits are lower by ~20 K compared with the results published in this work, but overall our $N$ and $T$ maps have a similar appearance and are consistent in their variations. The differences may be due to the fact that our fitted maps are at a slightly coarser resolution from smoothing to Herschel 350 μm maps, as well as a lack of longer-wavelength data to better constrain the temperature. Nonetheless, since our maps follow the same basic morphology as the ones presented in D. T. Chuss et al. (2019), we are sufficiently confident in the trends seen using only four wavelengths to utilize our fits. We find a maximum H$_2$ column density of $4.6 \times 10^{23}$ cm$^{-2}$, with a temperature ranging from ~30 to 70 K.

From our fitting, we find that M17 has a maximum column density of $4.1 \times 10^{23}$ cm$^{-2}$. This is similar to the result from Herschel graybody fitting done by T. D. Hoang et al. (2022), who used a coarser resolution. Notably, the temperature of M17 is influenced by the heating from O stars that lie just outside of the region we mapped (R. Chini et al. 1980; D. Lemke & A. W. Harris 1981; J. L. Dotson 1996). This effect can be seen in our temperature fits (see the right-middle panel of Figure 2), where the maximum temperature occurs on the east side of the map at ~125 K. The temperature of the dense regions of M17 is ~35 K. This is very similar to the temperature (~39 K) found by T. D. Hoang et al. (2022) in this region. The overall differences seen in the fitting could also be due to the inclusion of Herschel 500 μm data by these authors.

In W3 we obtain a maximum H$_2$ column density of $1.9 \times 10^{23}$ cm$^{-2}$. This is consistent with the value found from A. Rivera-Ingraham et al. (2013) in the same region. A. Rivera-Ingraham et al. (2013) also fit the dust temperature of W3 using multiple Herschel bands and found a maximum temperature of ~32 K in the cloud. Their fits for temperature are systematically lower than ours, with our peak temperature reaching ~54 K. There are several possible causes for the difference. First, these authors fit the temperature using images that have a coarser pixel grading than ours (36″ compared to 25″). This could have the effect of smoothing out high temperatures over multiple pixels and ultimately lowering the calculated temperature. These fits were also made using a different value of $\beta$ (2), so the $T - \beta$ covariance (e.g., X. Dupac et al. 2001, 2003; R. Shetty et al. 2009) may be contributing to the difference. Additionally, and perhaps more importantly, these authors used longer-wavelength data to fit the temperature. These data are more likely to constrain the cold dust temperatures than shorter-wavelength data. The temperature gradients seen here, however, are similar to the ones in A. Rivera-Ingraham et al. (2013).

### 3.3. Polarization Analysis

The goal of this analysis is to ultimately use these polarization observations as a probe of the dust alignment physics. To do this, we first calculate the median polarization spectrum in each cloud, as shown in Figure 3. We normalize the polarization percentage in each pixel at 89 and 154 μm to the polarization percentage in the corresponding pixel in the 214 μm image and then take the median value across each cloud. The error bars in Figure 3 are the median absolute deviation (MAD) from the median normalized polarization at each band and are essentially measuring the spread in normalized polarization values we see for each cloud. We summarize our findings for both the median polarization percentage across each cloud and the median normalized polarization in Table 1.

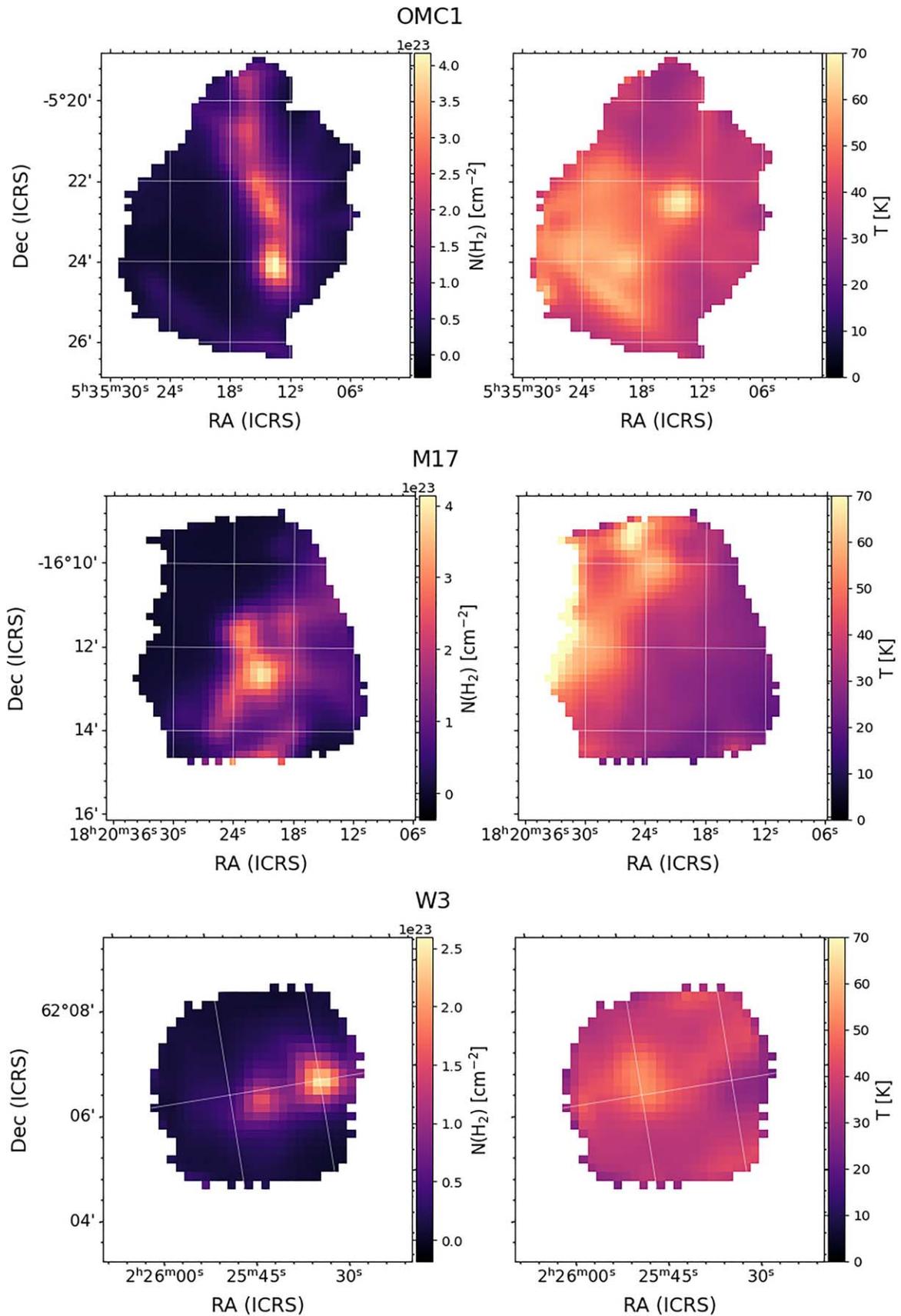

**Figure 2.** Column density (*N*; left panels) and temperature (*T*; right panels) of OMC-1 (top), M17 (middle), and W3 (bottom). These fits were made using the total intensity in the three HAWC+ bands presented here, as well as the Herschel 350 $\mu$m SPIRE maps. The fits are at the same resolution, 25″, as the Herschel data used.

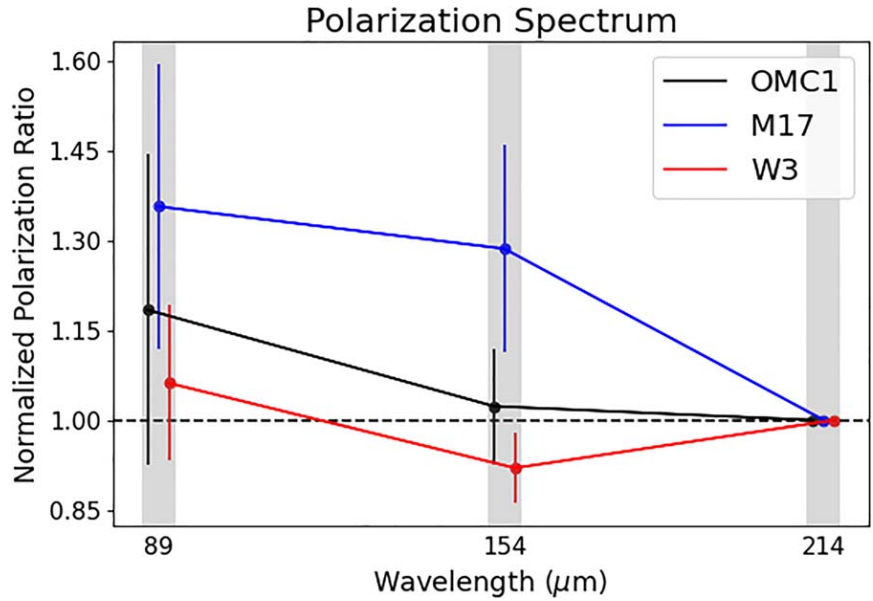

**Figure 3.** Median polarization spectrum for all three clouds. We normalize the polarization percentage at each wavelength to 214 μm and then calculate the median polarization across the normalized maps. Error bars show the absolute deviation from the median of the normalized polarization percentage. A dashed horizontal black line denotes an equal normalized polarization. The wavelength is offset for OMC-1 and W3 for ease in viewing the values of the points. Both OMC-1 and M17 exhibit a clearly falling spectrum, while W3 is flatter. OMC-1 shows the biggest drop between 89 and 154 μm, whereas in M17 the biggest drop is between 154 and 214 μm.

It can be seen from Figure 3 that all three clouds show a falling polarization spectrum, i.e., an overall trend of decreasing normalized polarization percentage with wavelength. Table 1 shows that the median polarization percentage in W3 is lowest at the short wavelengths and highest at the long wavelengths. This result, however, is likely skewed by the fact that there are fewer independent detections at 89 μm than at both 154 and 214 μm (see Figure 1). The 89 μm data are concentrated toward the densest regions, where clouds typically show smaller polarization percentage (see discussion surrounding Figure 4, below). Overall, the polarization values in Table 1 are highest for OMC-1 and lowest for W3. These cloud-to-cloud variations may be explained by their differing distances, due to beam dilution effects. However, W3 and M17 lie nearly at the same distance, so the smaller polarization values in W3 are unlikely to be caused by distance alone.

Figure 3 shows that in OMC-1 and W3 the largest change in normalized polarization percentage is between 89 and 154 μm, while M17 has the largest change between 154 and 214 μm. For the case of OMC-1, our results agree with the similar analysis carried out by J. M. Michail et al. (2021). For example, we find a decrease of 0.18 between 89 and 214 μm, while they find 0.19. There are some differences between their analysis and ours. They used a finer resolution (20″ vs. 25″) and implemented slightly different data cuts. In addition, J. M. Michail et al. (2021) include a polarization measurement at 53 μm and find that the spectrum is flat between 53 and 89 μm. Our results for M17 are the first far-IR polarization spectrum study for this cloud using SOFIA data. R. H. Hildebrand et al. (1999) studied this cloud using a combination of KAO and ground-based observations, showing that the polarization falls steadily from 60 to 100 to 350 μm. It is difficult to make quantitative comparisons between their work and ours, as their three wave bands are so much more widely separated than ours. Our results for W3 represent the first far-IR polarization spectrum study for this cloud. Here we see a spectrum that could possibly start to increase again at longer wavelengths. This is reminiscent of the "V" shape seen in the overall far-IR/submillimeter polarization spectrum of other clouds (e.g., J. E. Vaillancourt & B. C. Matthews 2012), but the minimum polarization value in the other clouds appears at longer wavelengths. The existence of the "V" shape in W3 is uncertain since the spectrum is nearly flat within the error bars between 154 and 214 μm.

To better understand polarization differences in each cloud, we compare the polarization percentage to the cloud's column density and temperature. In Figure 4 we show the polarization percentage for each SOFIA wavelength and cloud binned in $N$ and $T$ (see Figures 7 and 8 in the Appendix to see how the polarization ratio changes spatially and with N and T). We limit the bins in these parameters to have the same range for all clouds. Each environmental parameter is divided into 50 bins;

**Table 1**
Polarization Percentage Statistics for Each Cloud

| Cloud | 89 μm | MAD | 154 μm | MAD | 214 μm | MAD |
|---|---|---|---|---|---|---|
| Median Polarization Percentage ||||||||
| OMC-1 | 5.28 | 1.80 | 4.77 | 1.46 | 4.86 | 1.70 |
| M17 | 3.64 | 1.52 | 4.34 | 1.82 | 2.80 | 1.26 |
| W3 | 2.42 | 0.63 | 3.10 | 1.50 | 3.47 | 1.45 |
| Median Normalized polarization ($p_\lambda/p_{214}$) ||||||||
| OMC-1 | 1.18 | 0.26 | 1.02 | 0.10 | 1.0 | ⋯ |
| M17 | 1.36 | 0.24 | 1.29 | 0.17 | 1.0 | ⋯ |
| W3 | 1.06 | 0.13 | 0.92 | 0.06 | 1.0 | ⋯ |

**Note.** Comparisons of the median polarizations across the clouds and their respective median absolute deviation (MAD).

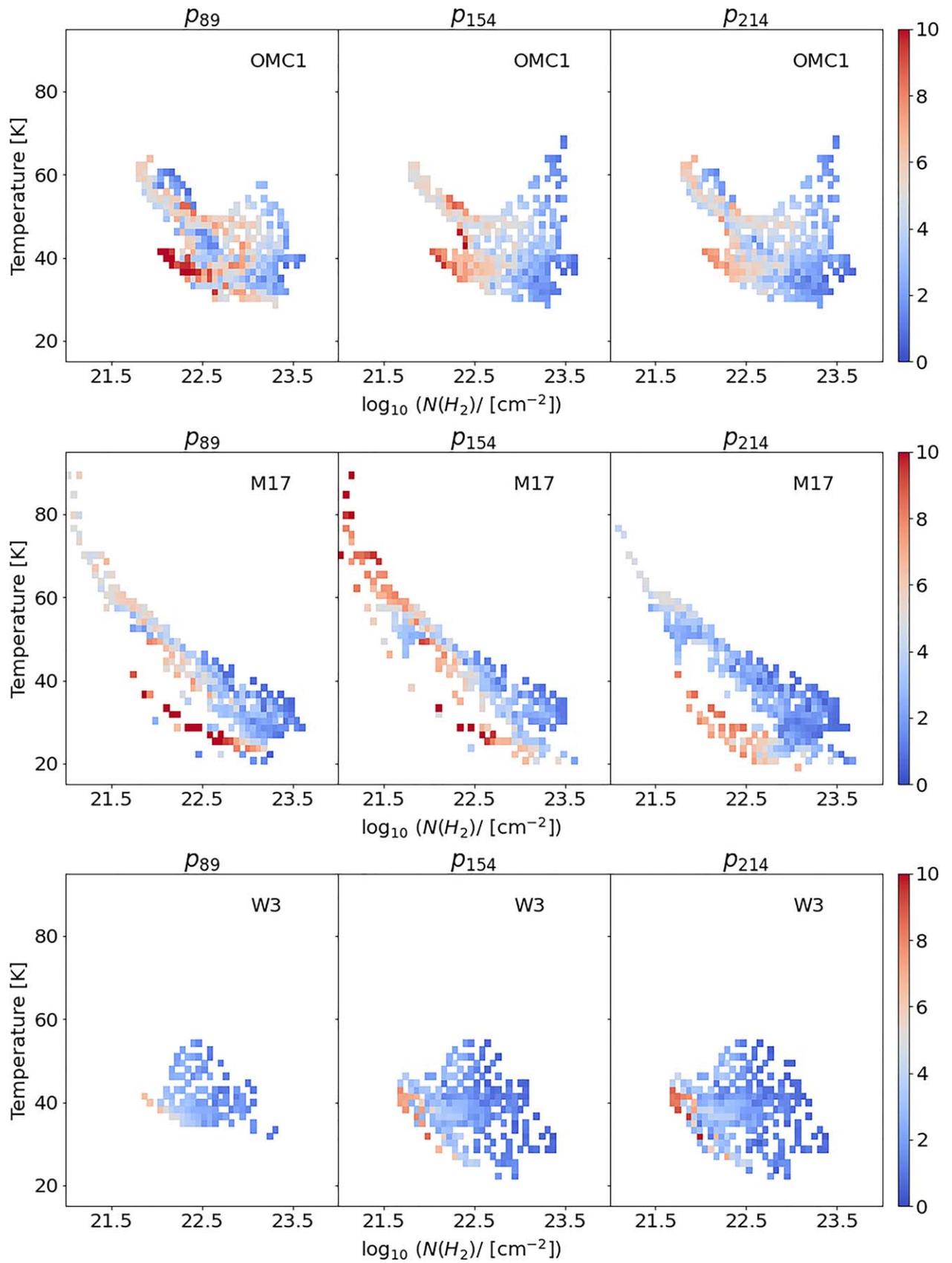

**Figure 4.** Color scale shows polarization percentage for OMC-1 (top), M17 (middle), and W3 (bottom), binned in $N$ ($x$-axis) and $T$ ($y$-axis). We keep the $N$ and $T$ bins the same for each cloud to see any differences. In each cloud we see lower polarization percentage (seen in blue) as column density increases.

column density is binned logarithmically, and temperature is binned linearly. Column density has a range of $10^{21}$–$10^{24}$ cm$^{-2}$, and temperature has a range of 15–95 K. After setting the bins, we compute the median value of the polarization percentage in each bin. We see in Figure 4 that the polarization percentage is lower in regions of higher column density across each cloud and wavelength. This depolarization at high column densities is discussed by B. C. Matthews et al. (2001) and has been seen in observations of star-forming regions over a range of spatial scales (e.g., L. M. Fissel et al. 2016; E. G. Cox et al. 2018).

Ultimately we are interested in how the polarization changes with wavelength in the star-forming regions. To quantify this, we follow previous authors (e.g., N. N. Gandilo et al. 2016; J. M. Michail et al. 2021) in fitting the polarization spectrum at each pixel to the equation

$$p(\lambda)/p(\lambda_0) = a_l(b_l[\lambda - \lambda_0] + 1). \quad (8)$$

In this equation, $\lambda_0 = 214\,\mu$m is the wavelength at which we normalize our polarization data. We use the parameter $b_l$ of this fit to determine whether the spectrum is rising or falling. The actual slope of the spectrum is $a_l \times b_l$; however, $a_l$ is $\sim$1, so we use $b_l$ as the important value to track throughout the paper and refer to $b_l$ as the slope. We use the same bins for $N$ and $T$ as computed above and find the median value of $b_l$ in each bin. This is shown in Figure 5. In this figure we see that OMC-1 and M17 both show a trend of a negative slope in regions corresponding to lower temperature. W3 exhibits a clear trend with a more positive slope in regions of lower column density and a negative slope at higher column densities.

To further investigate the extent to which $b_l$ is sensitive to changes in column density and temperature for each cloud, we compare $b_l$ to each parameter, shown in Figure 6. Here we plot the values of $b_l$ (red points) for each cloud versus $T$ (left) and $N$ (right). To look for trends, we again bin $N$ and $T$; however, these bins do not have the same bounds as discussed previously. We group each of $N$ and $T$ into 10 distinct bins that are chosen separately for each cloud. Again, $N$ is binned logarithmically and $T$ is binned linearly. We then find the median value of $b_l$ in each bin, shown in the black points of Figure 6, and the median average deviation within the bin, shown as error bars. In our discussion of Figure 3 above, we noted that the polarization spectrum of W3 has a much gentler downward slope compared to what we see in the other two clouds. However, in Figure 6 we can see that for the highest column densities W3 exhibits values of $b_l \sim -2.5$, similar to the medians found in the other two clouds.

We next perform statistical tests on the median values of $b_l$ and present these results in Table 2. We use the Pearson $r$ coefficient to constrain whether $b_l$ is correlated with either $N$, $T$, or both. We also use the two-tailed probability $p_{tt}$ to understand the likelihood of our results being random as a test of the significance of these results. We find that the polarization spectrum slope in M17 has a significant positive correlation ($3.3\sigma$) with temperature and that the slope in W3 has a significant negative correlation ($3.6\sigma$) with column density. We find a marginal ($2.9\sigma$) positive correlation with temperature in OMC-1 and no significant correlation to column density. In their analysis of OMC-1, J. M. Michail et al. (2021) used the same statistical tests and found a significant positive correlation with temperature ($3.9\sigma$) and no significant correlation to column density ($0.2\sigma$). These results are similar to ours. Their

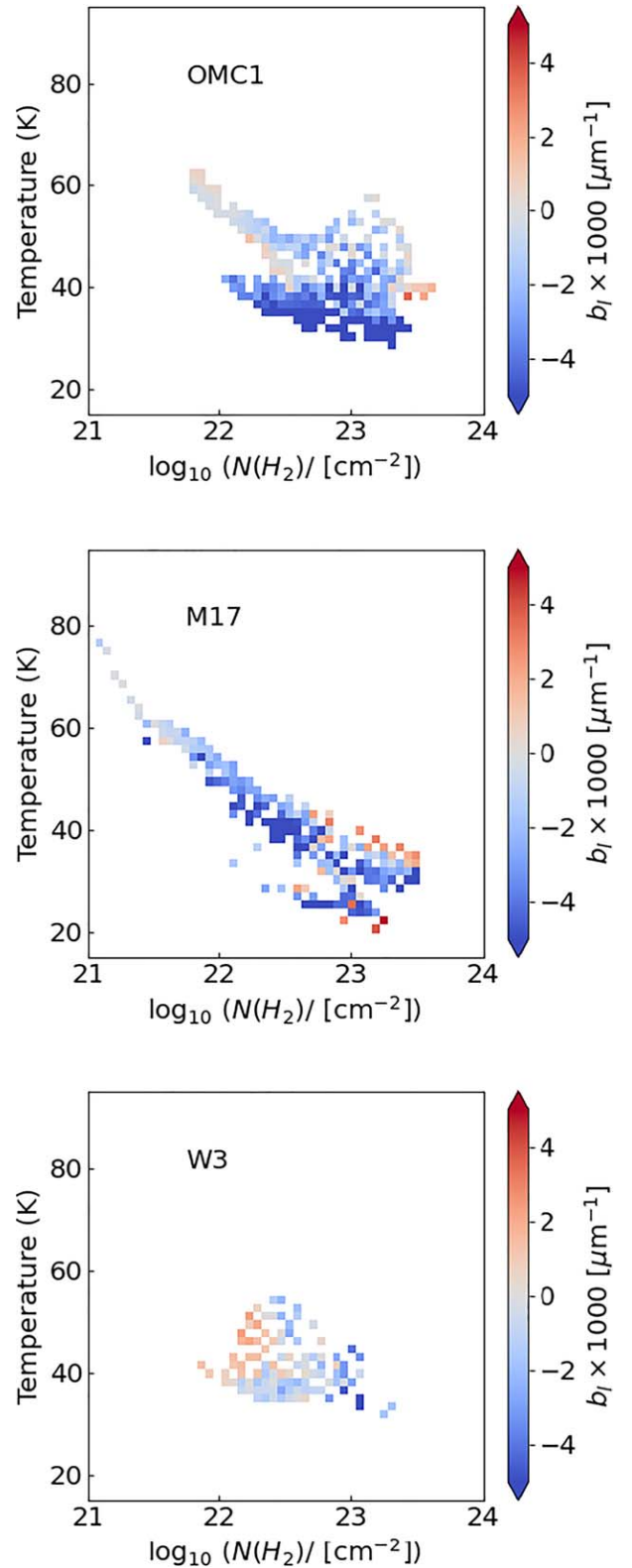

**Figure 5.** Fitted polarization slope ($b_l$) binned in $N$ and $T$ for each cloud (OMC-1 on top, M17 in the middle, and W3 on bottom), for the same bins used in Figure 4. It can be seen here that each cloud has an overall falling spectrum (shown in blue). OMC-1 and M17 tend to have lower values of $b_l$ at lower temperatures for a given column density bin.

$r$ values are also similar to ours in being near +1.0 for temperature and near 0.0 for column density. Specifically, their values are 0.93 and 0.08, respectively. The Appendix shows

**Table 2**
Correlations of $b_l$ with $T$ and $N$

| Parameter | Cloud | Pearson $r$ Coefficient $r$ | Probability $p_{tt}$ | Sigma $\sigma$ |
|---|---|---|---|---|
| $T$ | OMC-1 | 0.811 | $4.37 \times 10^{-4}$ | 2.9 |
|   | M17 | 0.877 | $8.66 \times 10^{-4}$ | 3.3 |
|   | W3 | 0.484 | $1.57 \times 10^{-1}$ | 1.4 |
| $N$ | OMC-1 | −0.283 | $4.28 \times 10^{-1}$ | 0.8 |
|   | M17 | −0.711 | $2.11 \times 10^{-2}$ | 2.3 |
|   | W3 | −0.901 | $3.61 \times 10^{-4}$ | 3.5 |

**Note.** Correlation statistics computed for medians of binned data.

additional visualizations of the polarization spectrum in our sample of clouds and how it varies with $N$ and $T$.

## 4. Discussion

### 4.1. Polarization Spectrum Shape

We find a falling polarization spectrum in each of the three clouds studied, although for W3 the overall spectrum falls by only a factor of 1.06, with sharply falling spectra confined to the highest column densities (see Section 3.3). Here we turn to the question of what mechanism(s) are responsible for these falling spectra. In Section 1 we noted that one such mechanism is the HCE, which invokes correlations between temperature and grain alignment properties along the sight line to explain the falling spectra. However, there a number of alternate explanations that we also consider here.

We first briefly discuss polarization mechanisms not involving magnetic alignment of dust grains. These mechanisms are typically observed on protostellar disk scales (∼100 au). On these scales grains may not have had enough time to align with the magnetic field (R. Tazaki et al. 2017), and polarized emission from self-scattering (A. Kataoka et al. 2015; E. G. Cox et al. 2018) or radiative alignment (R. Tazaki et al. 2017; H. Yang et al. 2019) may be the dominant mechanism. While these mechanisms can produce a falling polarization spectrum, they seem unlikely to be relevant to the results shown here, as the smallest resolution element we probe is in OMC-1 at ∼10,000 au scales.

All other effects that produce a falling polarization spectrum discussed here are based on dust grains that are preferentially aligned with their angular momentum parallel to the local magnetic field and emit polarized light orthogonal to their axis of rotation (e.g., A. Lazarian 2007; B. G. Andersson et al. 2015). The first such mechanism we consider is that of grain composition effects acting alone, i.e., with all grains in the cloud exposed to the same radiation field. Some theoretical modeling work in this area shows rising polarization spectra in the 89–214 $\mu$m range (B. T. Draine & A. A. Fraisse 2009; V. Guillet et al. 2018), while other such work shows that falling spectra can also be produced (H. Lee et al. 2020; single-phase models of L. N. Tram et al. 2024). However, in these models the polarization fraction does not fall fast enough between 89 and 214 $\mu$m to explain our observations. Specifically, the most sharply falling spectra have ratios of 89–214 $\mu$m polarization below ∼1.16, while we see median values of 1.18 and 1.36 in OMC-1 and M17, respectively (see Table 1). To explain the sharply falling spectra we have observed, in the remainder of this section we consider scenarios that rely on separate warm and cool regions along the line of sight.

We consider whether the falling polarization spectrum we observe could be caused by magnetic field tangling. This effect may occur when there are accreting protostars that are pulling in the surrounding material. Due to flux frozen magnetic field lines in the material, this infall causes the field closer to the cold, dense regions (probed by longer wavelengths) to be complicated within a resolution element and essentially cancels out the overall polarization. This does not mean that the magnetic field loses its strength in dense, cold regions owing to effects such as ambipolar diffusion. The drop in polarization percentage seen as the wavelength increases is a geometrical phenomenon caused by the infalling material. In this scenario, we would expect to see two observational signatures: (1) disordered magnetic field orientations near the dense regions forming protostars, and (2) a falling spectrum also near these regions. While we observe a fall in polarization percentage near the dense regions in some of our sample (see Figure 1), we do not find that regions showing a falling spectrum are spatially correlated with regions showing field disorder (compare Figure 9 in the Appendix with Figure 1). For this reason it does not seem likely that field tangling due to infalling material is the dominant cause of the falling polarization spectrum observed in our sample.

Another mechanism that can affect the polarization spectrum is RAT disruption (RAT-D). This effect is within the RAT paradigm and relies on two main components: (1) grains larger than $a \gtrsim 0.1$ $\mu$m (T. Hoang 2020), and (2) a sufficiently strong radiation field to "disrupt" dust grains (i.e., the radiation field destroys the dust grains; T. Hoang et al. 2019). Using SOFIA 154 $\mu$m polarization data, T. D. Hoang et al. (2022) argue that the data provide evidence for RAT-Ds in the northern region of M17 (not studied by us). The authors of this study suggest that this is seen in the polarization percentage data that is rapidly decreasing in regions of high temperature, where the column density and polarization angle dispersion are both low. In our data, we find that the regions of M17 that have high $T$ and low $N$ have systematically higher polarization percentage measurements at 154 $\mu$m (see middle panel of Figure 4). T. D. Hoang et al. (2022) do not find evidence for RAT-Ds in the southern region of M17 that we are probing, but we can look for RAT-Ds in this region using our own data. While we cannot directly compare their study with our analysis, due to spatial resolution differences, when we examine regions of high temperature in M17 across all three wavelengths, we find results consistent with what T. D. Hoang et al. (2022) found in this region—there is no evidence for RAT-Ds in the southern region of M17 (M17-SW). When we analyze our entire sample for evidence of RAT-Ds, we do not find regions of high polarization percentage in regions of high temperatures, though OMC-1 (see top panel of Figure 4) does show regions of moderate polarization percentage interspersed with low polarization percentage. Using these results, we do not find strong evidence for RAT-Ds in our sample.

The next mechanism we consider for producing a falling polarization spectrum is through HCE (R. H. Hildebrand et al. 1999; J. M. Michail et al. 2021). This effect relies on the star-forming cloud having temperature variations along the line of sight, with warmer regions having better-aligned grains and cooler regions having grains that are less well aligned. The temperature variations are presumed to arise because of shielding, with warm regions preferentially exposed to short-wavelength (i.e., UV/optical/near-IR) radiation from the interstellar radiation field and/or from embedded sources in the cloud itself. The RAT mechanism naturally explains the HCE, with the same radiation that warms grains also causing magnetic alignment. D. Seifried et al. (2023) present a semianalytic dust grain model that includes two temperature components—separate hot and cool regions along the line of sight—and is able to explain the falling far-IR polarization spectra observed using KAO. In this model the silicate grains are well aligned to the magnetic field and the carbon grains are not. D. Seifried et al. (2023) obtain their best fit to observations when they include both the HCE effects and a novel mechanism: UV-induced carbon grain destruction in the hot component.

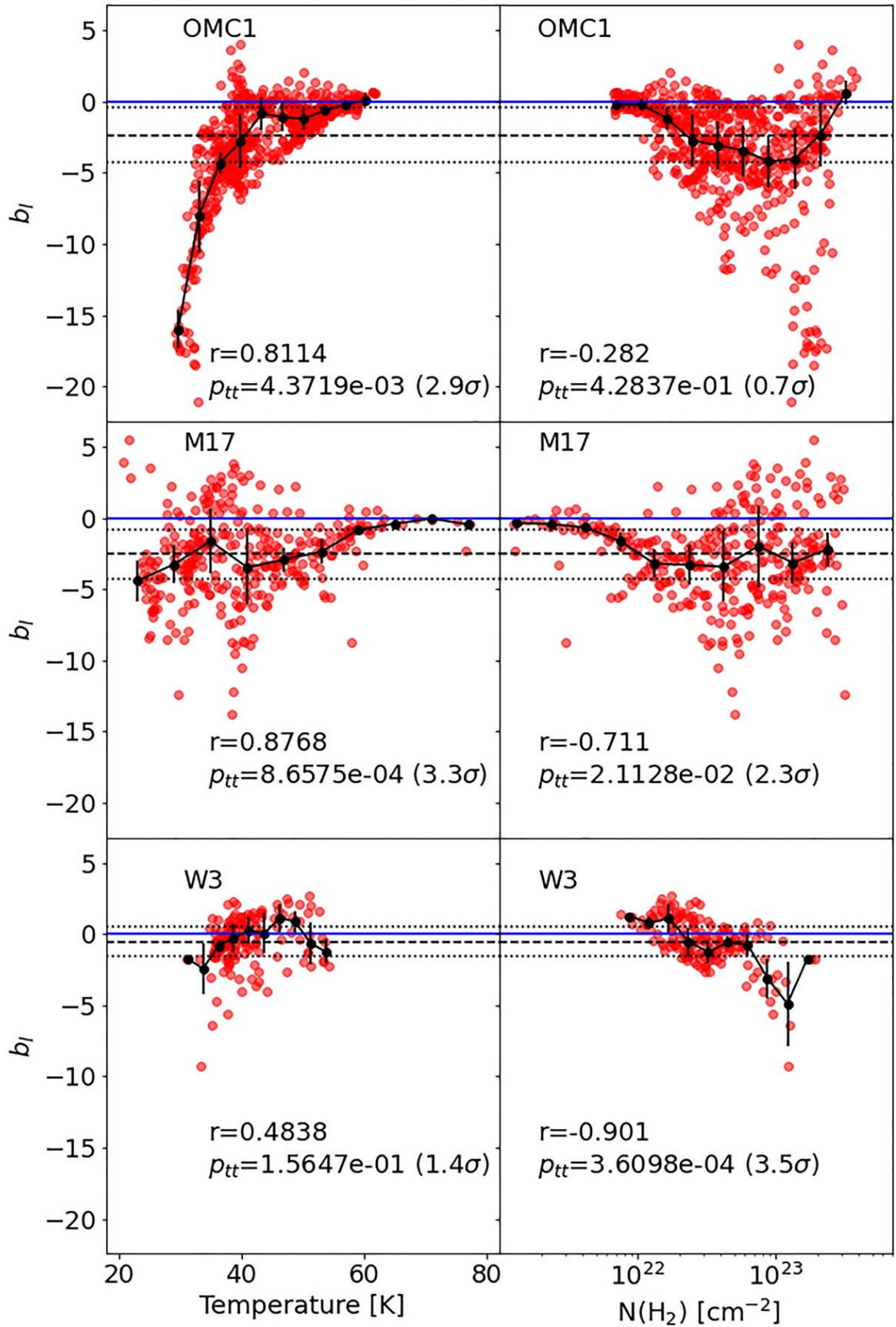

**Figure 6.** Values of $b_l$ for each cloud—OMC-1 (top), M17 (middle), and W3 (bottom)—plotted vs. temperature (left) and column density (right). Red points denote the value of $b_l$ for each point fitted, while the black points are median values for bins of $b_l$. To create these bins, we evenly divided each of $N$ and $T$ into 10 bins, using logarithmic spacing for $N$. Error bars denote the MAD in $b_l$ corresponding to each bin. Black dashed lines denote the median values of $b_l$ for each cloud, and dotted black lines show plus/minus the MAD levels of $b_l$ for the entire sample. A solid blue line is plotted at $b_l = 0$ to better compare $b_l$ values.

L. N. Tram et al. (2024) use realistic physical grain models to understand the polarization spectrum in OMC-1 and find that their two-phase model (i.e., two kinds of regions with different radiation strengths) is the best fit to the data. In particular, by changing different free parameters of this two-phase model, they are able to fit various spectrum shapes observed in OMC-1, including the fall in polarization seen at far-IR and the rise in polarization observed at longer wavelengths. The L. N. Tram et al. (2024) results for this model show a sharp decline in the far-IR region of the polarization spectrum, suggesting that it could also be used to model the other two clouds presented in this paper. Since this model uses two distinct radiation fields, it is not incompatible with the aforementioned HCE mechanism, as shielding could provide a way to obtain varying radiation field strengths along the line of sight.

Our analysis of the polarization spectrum implies the need for multiple temperatures within the star-forming cloud to produce the falling shape observed. In this section, we explored several mechanisms that rely on a temperature variation to better understand the dust physics in OMC-1, M17, and W3. Based on the arguments presented here, we favor HCE and/or carbon grain destruction scenario(s) to explain the falling far-IR polarization spectrum seen from our data, although additional physical effects incorporated into the models of L. N. Tram et al. (2024) should also be considered.

### 4.2. Column Density Dependence

We analyzed the polarization spectrum slope to determine whether it depends on either temperature or column density, or both. This is because, as discussed above, this dependence could be indicative of these environmental properties affecting the dust grains. In our sample of three clouds, we find that only one, W3, has a significant anticorrelation with N. In an analysis using 89 and 154 $\mu$m HAWC+ polarization, F. P. Santos et al. (2019) found a similar result in the $\rho$ Oph A star-forming region. This result is consistent with grain alignment becoming less efficient in the shielded, dense regions of the cloud, which is likely the effect we are probing in W3. More generally, when we examine the diffuse regions of our three clouds, we find that $b_l$ approaches zero (i.e., the spectrum flattens) below some critical column density in each of the three clouds (see right panels of Figure 6). This result implies that without a sufficient quantity of dust a falling polarization spectrum cannot arise.

To better understand why there is a critical column density below which the polarization spectrum is flat, we compare our values for the critical column density with results from other studies investigating dust grain alignment. In particular, D. C. B. Whittet et al. (2008) and T. J. Jones et al. (2015) find that when optical, IR, and submillimeter polarization percentage measurements are compared to extinction ($A_v$), there is a critical value of $A_v$ above which trends diverge, and this is attributed to a loss of grain alignment. T. J. Jones et al. (2015) use a toy model to find the value of $A_v$ at which grain alignment breaks down. Including turbulence in their model, T. J. Jones et al. (2015) find a break in the polarization trends at $A_v = 20$. Using interstellar extinction curves from G. H. Rieke & M. J. Lebofsky (1985), they find $A_v/A_\lambda = 20$ at $\lambda = 3.7$ $\mu$m, implying that the break occurs at $A_\lambda = 1$ for $\lambda = 3.7$ $\mu$m. This means that the polarization break corresponds to where the core becomes optically thick to radiation having wavelength $\lambda \lesssim 3.7$ $\mu$m. D. C. B. Whittet et al. (2008) also study where the trends in polarization with $A_v$ break down in their data and find a value closer to $A_v = 10$, though their model does not include turbulence.

We use these two results to better understand the trends we find in the polarization spectrum slope ($b_l$) compared to $N$, specifically the critical column density where $b_l$ approaches zero. While the exact value of the critical column density is cloud dependent, it is between $\sim 6 \times 10^{21}$ cm$^{-2}$ and $2 \times 10^{22}$ cm$^{-2}$, corresponding to an $A_v$ between 6 and 20 (using the conversion of $N(H_2) \sim 10^{21}$ at $A_v = 1$; R. C. Bohlin et al. 1978). This analysis yields values within order-of-magnitude limits of both the D. C. B. Whittet et al. (2008) and T. J. Jones et al. (2015) results and suggests that shielding of near-IR radiation at this critical column density influences the alignment of dust grains to the magnetic field in our three clouds. Since near-IR polarization and far-IR polarization are the result of different mechanisms, it is not certain that the far-IR polarization can be a probe of effects seen in near-IR polarimetry. This is especially true if grain growth has occurred inside the cloud (T. Hoang et al. 2021). However, if the far-IR polarization spectrum can be a reliable probe of some of the near-IR effects, then our results may further suggest that this near-IR shielding effect is necessary to produce the inferred temperature gradient discussed in Section 4.1. Because HCE requires shielding of near-IR radiation while carbon grain destruction does not, the column density dependence seen in our data would then suggest that HCE is at least partly responsible for the observed falling far-IR polarization spectrum.

### 5. Conclusions

In this paper we analyzed the far-IR polarization spectrum in three high-mass star-forming clouds to better understand the grain alignment physics in such clouds. From this analysis we find the following:

1. We replicate the general findings of J. M. Michail et al. (2021) for OMC-1 using only four observing wavelengths for $N$, $T$ fits, rather than 11. These include a falling polarization spectrum and a positive correlation between polarization spectrum slope and temperature.
2. Each cloud exhibits an overall falling spectrum using a large number of sight lines. All three clouds are nearby, yet there is significant variability seen in the data. Overall, the polarization in M17 falls by $\sim 35\%$, that in OMC-1 falls by $\sim 20\%$, and that in W3 only falls by $\sim 5\%$. The reason for this spread in data remains unknown.
3. There is a critical column density ($N \lesssim 10^{22}$ cm$^{-2}$) below which the falling polarization spectrum is not observed. This corresponds to an extinction, $A_v$, of $\sim 10$.

4. This value for the critical column density suggests that shielding of near-IR radiation is important for producing a falling polarization spectrum, perhaps by limiting the photons available for grain alignment in the shielded (cold and dense) interiors of high column density sight lines that emit at preferentially longer wavelengths.
5. W3 shows mostly lower polarization than the other two star-forming regions. This is likely not all due to distance effects.

Polarized emission can be a powerful tool to probe the dust physics in star-forming clouds. Examining the polarization spectrum in three clouds using three far-IR wavelengths, we found similarities across each environment. To fully utilize this tool to constrain the dusty environment in which stars form, however, will require more sensitive, multiwavelength polarization observations.

### Acknowledgments

This material is based on work supported by the National Science Foundation MPS-Ascend Postdoctoral Research Fellowship under grant No. 2213275. Any opinions, findings, and conclusions or recommendations expressed in this material are those of the author(s) and do not necessarily reflect the views of the National Science Foundation.

This work was based (in part) on observations made with the NASA/DLR Stratospheric Observatory for Infrared Astronomy (SOFIA). SOFIA is jointly operated by the Universities Space Research Association, Inc. (USRA), under NASA contract NNA17BF53C, and the Deutsches SOFIA Institut (DSI) under DLR contract 50 OK 2002 to the University of Stuttgart. Financial support for this work was provided by NASA through award No. 09-0535 issued by USRA. Z.-Y.L. is supported in part by NASA grant No. 80NSSC20K0533 and NSF AST-2307199. J.M.M. is supported by an NSF Astronomy and Astrophysics Postdoctoral Fellowship under award AST-2401752.

PACS has been developed by a consortium of institutes led by MPE (Germany) and including UVIE (Austria); KU Leuven, CSL, IMEC (Belgium); CEA, LAM (France); MPIA (Germany); INAF-IFSI/OAA/OAP/OAT, LENS, SISSA (Italy); and IAC (Spain). This development has been supported by the funding agencies BMVIT (Austria), ESA-PRODEX (Belgium), CEA/CNES (France), DLR (Germany), ASI/INAF (Italy), and CICYT/MCYT (Spain). This publication makes use of PACS data products. SPIRE has been developed by a consortium of institutes led by Cardiff University (UK) and including Univ. Lethbridge (Canada); NAOC (China); CEA, LAM (France); IFSI, Univ. Padua (Italy); IAC (Spain); Stockholm Observatory (Sweden); Imperial College London, RAL, UCL-MSSL, UKATC, Univ. Sussex (UK); and Caltech, JPL, NHSC, Univ. Colorado (USA). This development has been supported by national funding agencies: CSA (Canada); NAOC (China); CEA, CNES, CNRS (France); ASI (Italy); MCINN (Spain); SNSB (Sweden); STFC, UKSA (UK); and NASA (USA). This publication makes use of SPIRE data products (NHSC 2020a, 2020b).

*Facility:* SOFIA (HAWC+), Herschel(SPIRE), Herschel(PACS).

*Software*: python, numpy (C. R. Harris et al. 2020), matplotlib (J. D. Hunter 2007), scipy (P. Virtanen et al. 2020), astropy (Astropy Collaboration et al. 2013; A. M. Price-Whelan et al. 2018), aplpy T. Robitaille & E. Bressert (2012); T. Robitaille (2019).

**Appendix**

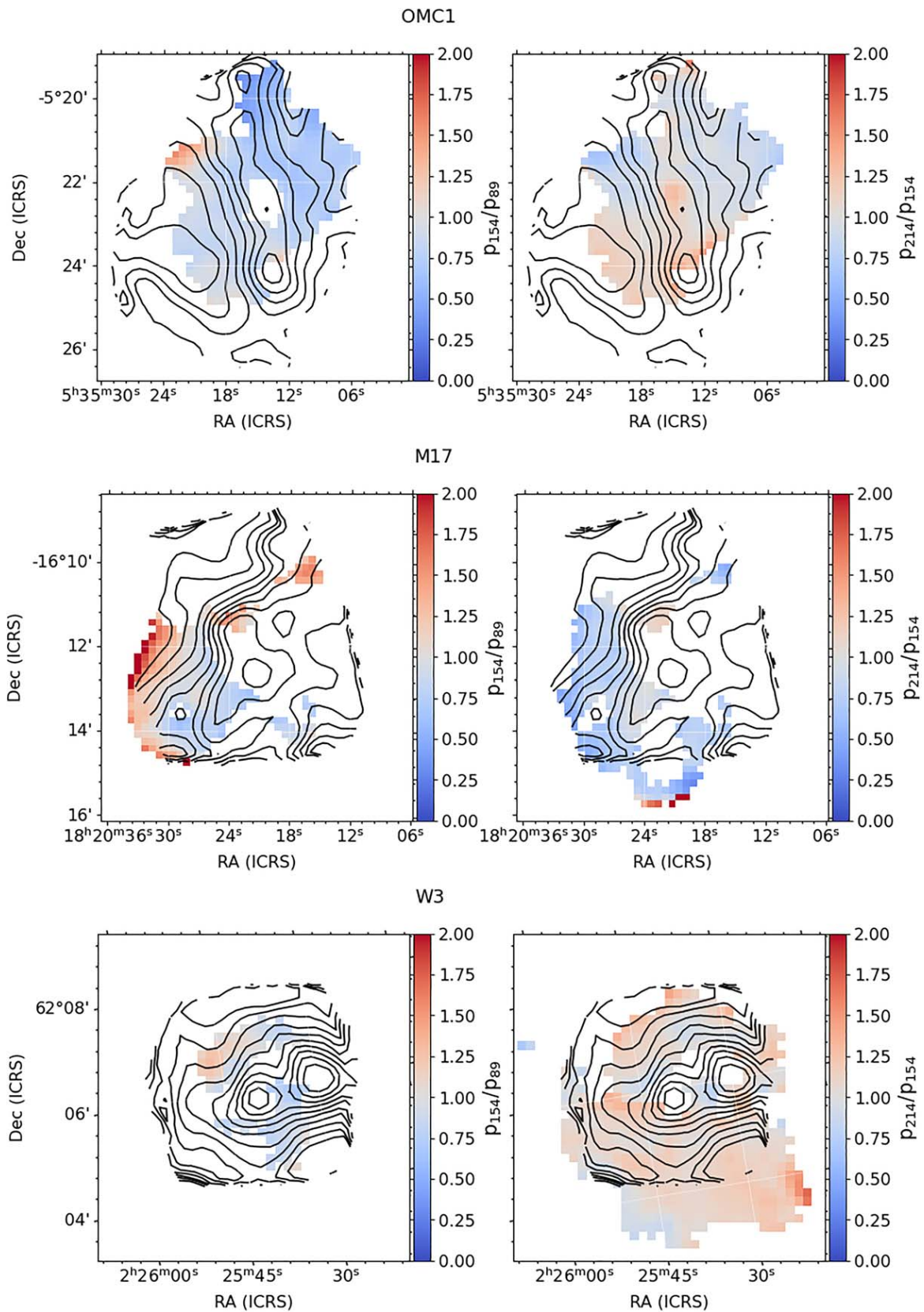

**Figure 7.** Maps of polarization ratios $p_{154}/p_{89}$ (left) and $p_{214}/p_{154}$ (right) for OMC-1 (top), M17 (middle), and W3 (bottom), spatially. Black contours show $N$ in 10 logarithmic spacings.

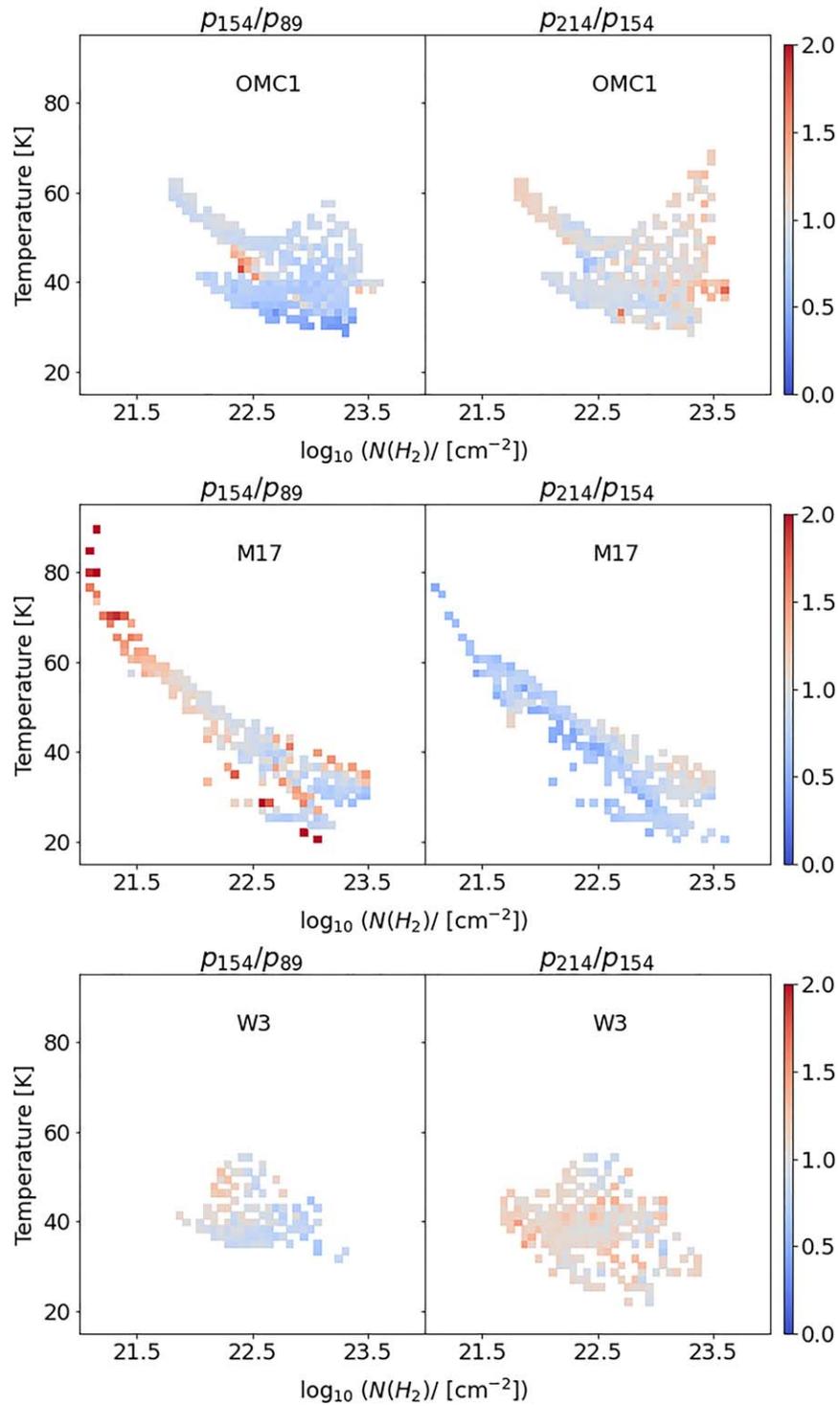

**Figure 8.** Polarization ratios $p_{154}/p_{89}$ (left) and $p_{214}/p_{154}$ (right) for each cloud binned using the same $N$ and $T$ bins as in Figure 4.

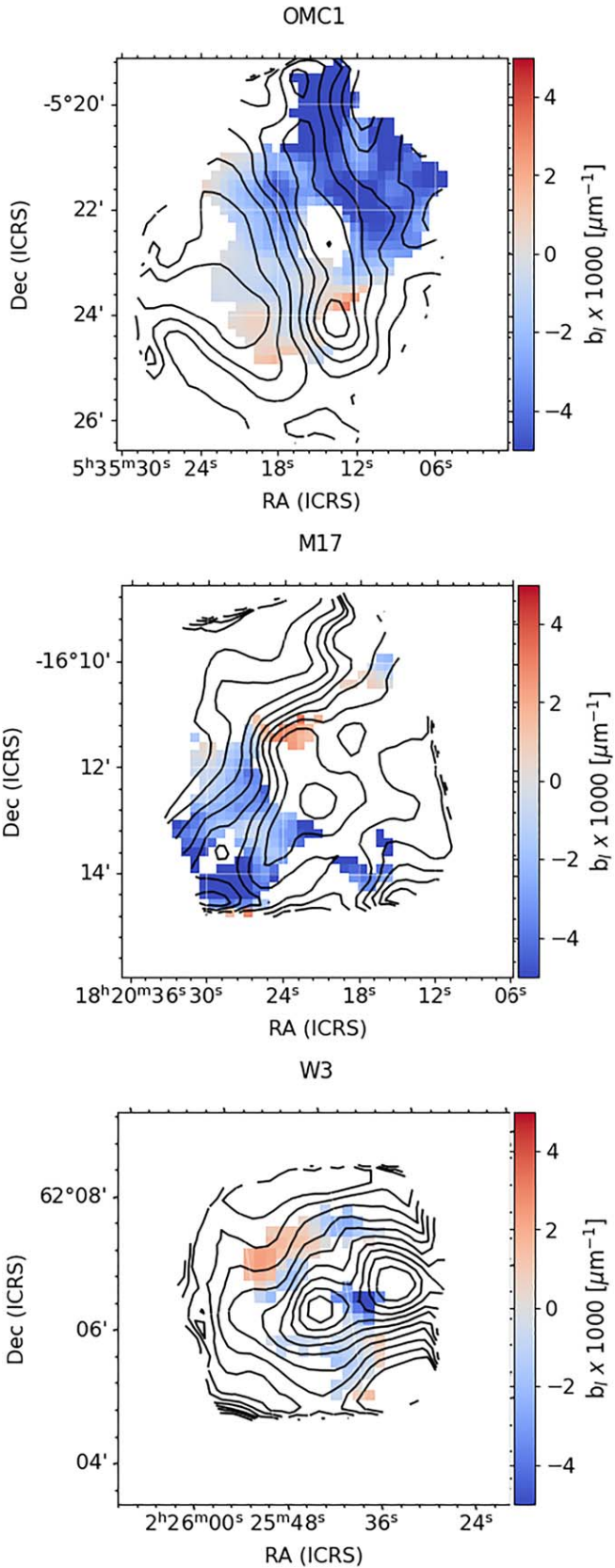

**Figure 9.** Parameter $b_l$ shown spatially across the cloud for OMC-1 (top), M17 (middle), and W3 (bottom). Black contours show $N$ in 10 logarithmic spacings.

## ORCID iDs

Erin G. Cox ● https://orcid.org/0000-0002-5216-8062
Giles Novak ● https://orcid.org/0000-0003-1288-2656
David T. Chuss ● https://orcid.org/0000-0003-0016-0533
Dennis Lee ● https://orcid.org/0000-0002-3455-1826
Kaitlyn Karpovich ● https://orcid.org/0009-0006-4830-163X
Joseph M. Michail ● https://orcid.org/0000-0003-3503-3446
Zhi-Yun Li ● https://orcid.org/0000-0002-7402-6487
Peter C. Ashton ● https://orcid.org/0009-0005-1673-0504